\documentclass[reprint, superscriptaddress, nobibnotes, amsmath, amssymb, aps, prl, floatfix, showkeys]{revtex4-2}
\usepackage{xr}
\usepackage{filecontents}
\usepackage{graphicx}
\usepackage{amsmath}
\usepackage{color}
\usepackage{xcolor}
\usepackage{subfiles}
\usepackage{float}
\usepackage{hyperref}
\hypersetup{colorlinks=true, linkcolor=blue, filecolor=blue, urlcolor=blue, citecolor=blue }
\newcommand{\overbar}[1]{\mkern 1.5mu\overline{\mkern-1.5mu#1\mkern-1.5mu}\mkern 1.5mu}

\begin{document}
\title{Cold self-lubrication of sliding ice}

\author{Achraf Atila}
\author{Sergey V. Sukhomlinov}
\author{Martin H. Müser}
\email{martin.mueser@mx.uni-saarland.de}
\affiliation{Department of Material Science and Engineering, Saarland University, Saarbr\"{u}cken, 66123, Germany}

\begin{abstract}
The low kinetic friction between ice and numerous counterbodies is commonly attributed to an interfacial water layer, which is believed to originate from pre-existing surface water or from melt water induced by high contact pressures or frictional heat. 
However, even the currently leading theory of frictional melting appears to defy direct experimental verification. 
Here we present molecular simulations of ice interfaces that reveal that ice surfaces liquefy without melting thermodynamically but predominantly by cold, displacement-driven amorphization.
Despite effective self-lubrication, very small ice friction is found to require water to slip past a hydrophobic counterface -- or an excess amount of water, produced by, e.g., extreme sliding velocities.
\end{abstract}
\date{\today}
\maketitle 

Skidding on ice or snow is a well-known phenomenon, often dreaded, sometimes loved.
The leading explanations for the low friction of ice and snow emphasize self-lubrication through water~\cite{Rosenberg2005PT, Dash2006RMP, Baran2022PNAS}.
However, the reason for the presence of water in sliding interfaces at sub-0$^\circ$ temperatures remains disagreed upon. 
Competing theories are pressure melting~\cite{Thompson1860PRSL},
surface melting~\cite{Faraday1859LEDPM, Slater2019NRC}, and friction-induced heating~\cite{Bowden1939PRSA}.
However, no single theory is conclusive~\cite{Lever2017JG}.
Pressure melting would require the true contact between a ski and the ice below it to be unreasonably
small to explain skiing at -20$^\circ$C~\cite{Bowden1939PRSA}.
Although the molecular mobility of sub-0$^\circ$C surface water correlates with ice friction~\cite{Weber2018JPCL}, 
the variation in friction coefficients with different counterbodies remains unexplained.
The arguably leading theory of melting by frictional heating must also be questioned:
warming of snow surfaces under a rotating slider at $-7^\circ$C temperature and 1~m/s sliding velocity could not be detected despite high temporospatial resolution~\cite{Lever2017JG}. 
Similarly, a contact-induced water film on ice produced at high sliding velocities of $v_0 = 5$~m/s~\cite{Baurle2006TL} did not exceed 0$^\circ$C.
Thus, either the main reason for the presence of water
in sliding ice interfaces varies from case to case, or some crucial ice-liquefaction mechanism has not been accounted for hitherto. 
One candidate mechanism could be related to Moras~\textit{et al.}'s hypothesis~\cite{Moras2018PRM} that sliding ice 
undergoes layer-by-layer amorphization as do diamond as well as silicon in incommensurate, self-mated contacts.

Determining the applicability of different mechanisms that produce interfacial water requires simultaneous analysis of interfacial stresses, temperature, and structure for diverse initial and boundary conditions in moving, buried interfaces.
This can be achieved by means of molecular dynamics (MD) simulations~\cite{Thompson2022CPC} when using force fields that accurately reproduce the relevant thermodynamic, dynamic, and mechanical properties, as the TIP4P/Ice potential does for water~\cite{Abascal2005JCP1, Vega2005JCP, Noya2007JPCC, Conde2008JCP, Baran2023JCP, RosuFinsen2023Science, IriarteCarretero2018PCCP}.
It allows us to effectively model sliding ice interfaces and investigate how sliding induces or maintains sub-0$^\circ$C interfacial water.
We start by simulating flat, incommensurate ice-ice interfaces.
Such calculations provide a lower bound for the friction between ice crystals as they disregard roughness-induced plasticity, plowing, snow compression, capillaries, and other processes that enhance energy dissipation.
To ascertain its load-bearing ability and the role of hydrophobicity, we also study ice sliding past corrugated counterfaces of different hydrophobicity. {Details on simulation setups and protocols are described in the supplementary materials (SM)~\cite{SuppMat}.}

\begin{figure*}[!ht]
\includegraphics[width=0.85\textwidth]{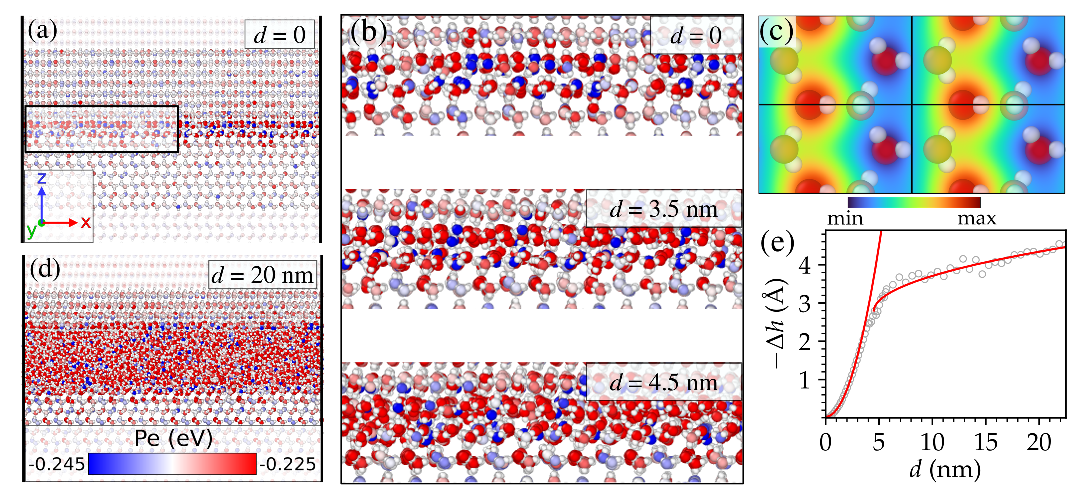}
\caption{\label{fig:static}
Snapshots of an incommensurate interface at $T = 10$~K (a) before the onset of sliding and (d) after sliding a distance of $d = 20$~nm at a velocity of $v_0 = 1$~m/s. 
Colors indicate the potential energy per molecule.
(b) Zooms of the interface before and during initial sliding. 
(c) Magnitude of the electrostatic field in a unit cell located one interlayer spacing above a free surface of an ideal crystal structure. 
Top-layer atomic positions are included.
(e) Negative change of total system height, $-\Delta h$, as a function of $d$.
Full lines reflect an initial harmonic response w.r.t. $d$ and shear-induced amorphization through a $\sqrt{d-d_0}$ dependence, where $d_0$ is the slid distance at the onset of plasticity. 
Visualizations of the atomic configurations are made using OVITO~\cite{Stukowski2009MSMSE}.
}
\end{figure*}

In the first set of simulations, two ice crystals are brought into contact at a temperature of $T = 10$~K with a small approach velocity $v_\perp$  until the normal force between their misaligned [0001] surfaces vanishes.
Localized zones of roughening, a few Angstroms wide, appear where the potential energy per molecule is lower than in the crystal, as evidenced in Fig.~\ref{fig:static}\,(a,b).
The low-energy zones arise when the dipoles of surface molecules are aligned with and thus attracted by the counterbody's electrostatic field, which is shown in Fig.~\ref{fig:static}(c).

Once sliding starts, these low-energy zones act like cold-welded spots causing plastic deformation in their vicinity, strengthening these surrounding areas but weakening the originally cold-welded site. 
In this way, old low-energy zones disappear, while new ones arise as sliding progresses.
Due to the open structure and low packing of hexagonal ice, dislocations are not needed for plastic deformation to occur during this process.
The instabilities that destroy crystalline order are local, whereby energy releases and associated temperature bursts are small. 
Similar dynamics were observed for other interfaces and another popular water potential, namely SPC/E~\cite{Berendsen1987JPC} (see examples shown in Fig.~\ref{fig:alternativeModels} in the SM~\cite{SuppMat}).
Since the [0001] surface is the most densely packed ice surface and the maximum misorientation of  $30^\circ$  provides the best possible condition for structural lubricity (SL), we can conclude that SL will not occur at other ice-ice interfaces either.
SL~\cite{Hirano1990PRB,Muser2004EPL,Dietzel2013PRL} refers to a state of small, Stokes-like friction, which is caused by the systematic cancellation of lateral forces when two atomically flat, incommensurate crystal surfaces slide past each other without invoking instabilities.

\begin{figure}[h!]
\centering
\includegraphics[width=\columnwidth]{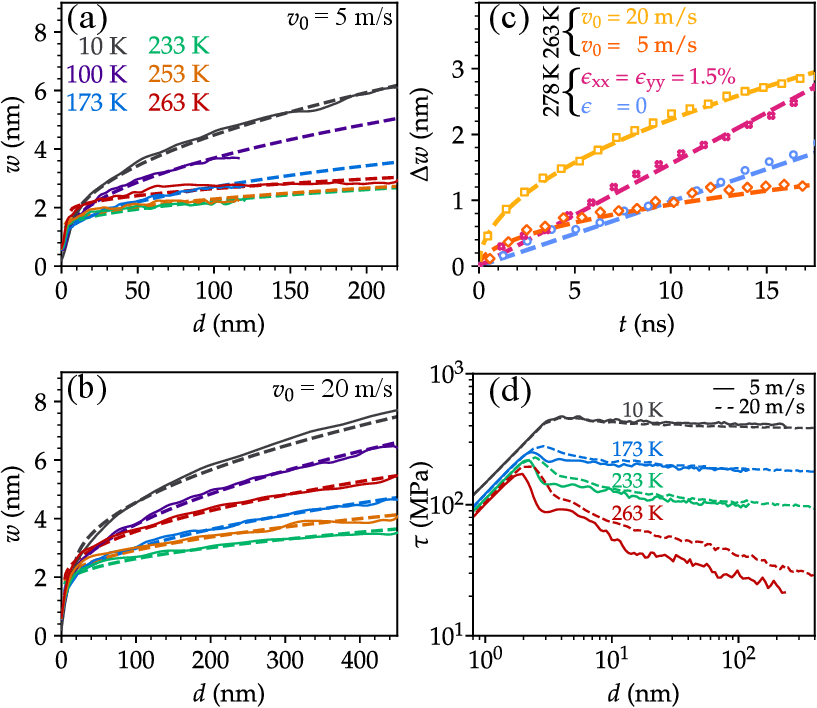}
\caption{\label{fig:widthStress} 
Width $w$ versus slid distance  $d$ at different temperatures $T$ for sliding velocity (a)  $v_0 = 5$~m/s and (b) $20$~m/s. Solid lines represent simulation data, while the dashed lines correspond to the square-root fits to the data.
(c) Increase of amorphous-layer width $\Delta w$ with time $t$ at $T = 263$~K for $v_0 = 5$~m/s (orange diamonds) and $v_0 = 20$~m/s (yellow squares) as well as at rest and $T = 278$~K: strain-free (blue circles) and under isotropic, in-plane strain (magenta crosses). Dashed lines are drawn to guide the eye.
(d) Shear stress $\tau$ as a function of slid distance $d$.
}
\end{figure}

The absence of SL in TIP4P/Ice interfaces does not imply that tetrahedral order is generally sufficient to suppress SL.
When using a popular mono-atomic model for molecular water (mW)~\cite{Molinero2008JPCB}, which favors tetrahedral order,
the shear stress is rather small for an atomically flat interface, i.e., $\tau \approx 6$~MPa, see Fig.~\ref{fig:mw_sliding} in the SM~\cite{SuppMat}. 
The mW model can create dry, low shear-stress contacts because particles lack internal degrees of freedom causing multi-stability and thus instabilities during sliding, which are produced by the orientational degrees of freedom of water molecules or by (re-) hybridization of carbon and silicon atoms.
Adding 200  particles to the 115~nm$^2$ large interface suppresses the incommensurability, thereby increasing the shear stress by a factor of almost twenty.
This trend of increased friction with added particles between atomically smooth surfaces is similar to that found in models of dense, atomically smooth surfaces with or without contamination~\cite{He1999S}.

Because shear stresses in \emph{dry}, incommensurate interfaces easily exceed 100~MPa even at -10$^\circ$C, 
friction between ice crystals can only be small given the presence of sufficiently thick pre-existing or tribo-disordered layers, whose structure [see, Fig.~\ref{fig:static}(d)] resembles that of sheared supercooled bulk water.
The claim of resemblance
is supported by comparisons of the pair distribution function (Fig.~\ref{fig:g2Combined} in the SM~\cite{SuppMat}), a three-body, mixed radial, angular distribution function~\cite{Sukhomlinov2020JCP} (Fig.~\ref{fig:g3Combined} in the SM~\cite{SuppMat}), and the concentration of five-coordinated water molecules (Fig.~\ref{fig:FFTemp} in the SM~\cite{SuppMat}).
The disordered zone has many five-coordinated molecules as regular and supercooled water, causing the liquid to be denser than the crystal. 
Thus, reductions in the separation between the two outermost layers $\Delta h$ allow the thickness or width $w$ of the amorphized zone to be determined.

Ice turns out to have the same proportionality of $-\Delta h_\textrm{p}$, see Fig.~\ref{fig:static}(e),
or $w_\textrm{p}$, see Fig.~\ref{fig:widthStress}(a,b) and movie~S1, to the square-root of the slid distance $\sqrt{d_\textrm{p}}$ as diamond and silicon~\cite{Pastewka2010NM,Moras2018PRM}.
$w_\textrm{p}$ was determined using the Chill+ algorithm~\cite{Nguyen2014JPCB} (details in the SM~\cite{SuppMat}) and the subscript p indicates that quantities are given relative to the point at which (substantial) amorphization sets in, which is also where the stiction peak is located, e.g., at $d \approx 3.5$~{nm} in Fig.~\ref{fig:static}(e).
The $w_\textrm{p} \propto \sqrt{d_\textrm{p}}$ relation indicates that the probability for a surface molecule to abandon its crystallographic position is linear in a distance increment $\Delta d_\textrm{p}$ but inversely proportional to $w_\textrm{p}$ and thus that amorphization is displacement-driven. 
Further evidence against thermal melting is provided in  Fig.~\ref{fig:widthStress}(c), which shows that shear-disordering the first nanometer of $-10^\circ$C cold ice with $v_0 = 5$~m/s takes roughly as long as it would take at rest to melt the same amount of ice of a fully thermostatted crystal superheated to $+5^\circ$C.
Yet, the temperature in the sliding system only rose to at most {$-5^\circ$C}, as shown in Fig.~\ref{fig:FFTemp}(b) in the SM~\cite{SuppMat}.
Tensile strains, which tend to be high at the trailing edge of sliding contacts~\cite{Muller2023PRL}, can be more important than heating since a 1.5\% isotropic, in-plane strain at $+5^\circ$C almost doubles the melting rate compared to the unstrained case, see  Fig.~\ref{fig:widthStress}(c).
Another argument against frictional heat as the main cause of shear-induced melting is that ice liquefies significantly faster at 10~K compared to -10$^\circ$C, a contrast made particularly evident when comparing movie S1 to movie S2.

While the amorphization coefficient $\alpha \equiv w_\textrm{p}^2/{d_\textrm{p}}$ changes non-monotonically with temperature---Fig.~\ref{fig:amorphizationrate} reveals a relative minimum occurs near $T=233$~K, i.e., slightly above the temperature separating the low- and high-density regimes of supercooled water~\cite{Gallo2016CR}---the shear stress decreases continuously with increasing temperature, as revealed in Fig.~\ref{fig:widthStress}(d).
This is because the effective viscosity, which is the ratio of shear stress to shear rate, has a strong temperature dependence~\cite{Baran2022PNAS,Baran2023JCP} as opposed to $\alpha$. 
Therefore, the reduction in viscosity due to frictional heating can be one reason why Bowden and Hughes~\cite{Bowden1939PRSA} observed insulating skis to have lower friction than heat-conducting skis, which, in their eyes, invalidated the viewpoints of pre-existing surface water and pressure melting to cause low ice friction.

Another explanation for Bowden and Hughes' observations arises from the possibility that the ice surface warms up more in reality than in our simulations.
This issue is important but also quite technical, which is why we address the specifics in the SM~\cite{SuppMat} and focus on the broader implications next. 
Once the near-surface regions, i.e., those where the thermostats act, exceed $-10^\circ$C, the recrystallization rate decreases rapidly as the temperature approaches the melting point~\cite{Xu2016PNAS,MonterodeHijes2019JCP}.
Therefore $w_\textrm{p}$ increases,  which reduces both shear rate and shear stress.
Data shown in Fig.~\ref{fig:non_stationary} and analyzed in the SM~\cite{SuppMat} suggests that this effect likely outweighs the impact of viscosity changes.
Thus, the Bowden and Hughes argument, together with modern theoretical estimates on `frictional heating and ice premelting'~\cite{Persson2015JCP}, may still hold, though not to the extent that liquefaction should be attributed to thermal melting.
In fact, ice friction can be low when the counterbody (e.g., metallic skates) conducts heat over 20 times better than ice, causing most frictional heat to transfer to the metal.

\begin{figure*}[ht]
\centering
\includegraphics[width=0.85\textwidth]{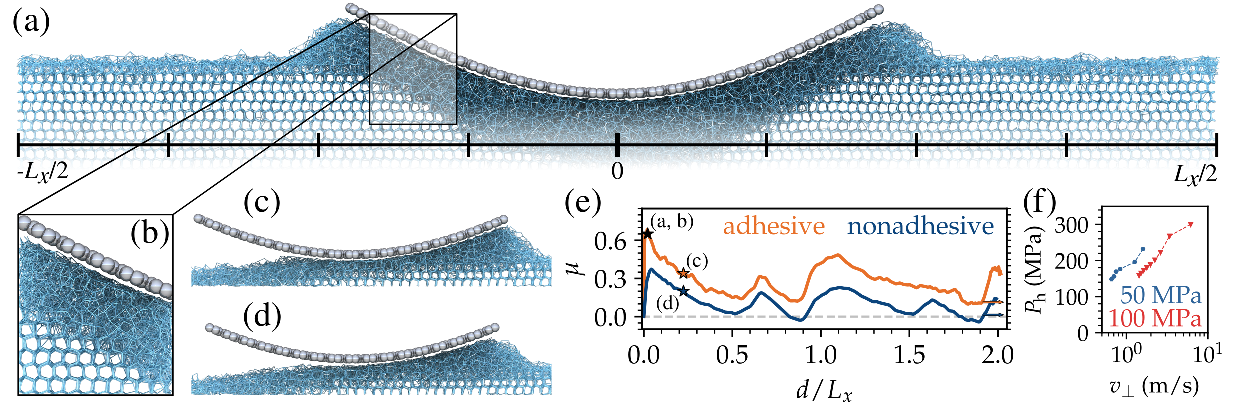}
\caption{\label{fig:indentation} 
(a) Molecular configuration of a slice of ice subjected to nanoindentation at $-10^\circ$C during the initial stage of sliding with $v_0 = 5$~m/s.   
(b) Zoom of the region highlighted by the box in (a).
Further zooms of other regions during sliding for adhesive [hydrophilic] (c) and nonadhesive [hydrophobic] (d) indenters.   
(e) Friction coefficient $\mu$ as a function of sliding distance $d$ normalized to the length $L_x=46$~nm of the simulation cell along the sliding direction.
Stars indicate the moment in time at which the snapshots shown in (a--d) were taken.
(f)~Penetration hardness ($P_\text{h}$) as a function of the indentation velocity using two normal loads.
In all snapshots, only O-O bonds are shown for clarity. 
Visualizations of the atomic configurations are made using OVITO~\cite{Stukowski2009MSMSE}.
}
\end{figure*}

In order to relate our results to experiments, shear stresses must be converted into friction coefficients $\mu$, defined as the ratio of shear force to normal load. 
This requires local water-film heights and normal contact pressures to be estimated, which is non-trivial, because they depend on contact-patch geometries, the squeeze-out dynamics of water, and the rate- and potentially scale-dependent ice plasticity or creep~\cite{Lahayne2016TL, Tada2023F, Persson2023JCP}.
However, once the approach velocity $v_\perp$ exceeds $0.1$~mm/s locally, our simulations reveal that normal pressures can far exceed the quasi-static penetration hardness $p_\textrm{H}$. This hardness is close to 10~MPa at temperatures where skiing and skating are most feasible~\cite{Liefferink2021PRX}, meaning close—but not too close—to melting. An example is $-10^\circ$C, the temperature we focus on in the following.

To generate realistic local stresses occurring during ice-asperity interactions without relying on rough continuum-mechanics calculations, we simulate a rigid, single-sinusoidal corrugated plate.
To this end, we first indent the tip into an initially flat ice surface at $-10^\circ$C (so that information on the stiction peak / static friction can be obtained) and then slide it. 
Parts of the final configuration of an indentation process with an adhesive rigid tip are depicted in Fig.~\ref{fig:indentation}(a,b).
It was produced by first applying a nominal normal pressure of $p_\textrm{n} = 100$~MPa for 0.6~ns, which was followed by a 40~ps relaxation at $p_\textrm{n} = 20$~MPa.
The high initial pressure creates an indentation mark of similar depth to that obtained with lower pressure over a time span far beyond the scope of molecular simulations. 

Corrugated indenters will sink into the ice while amorphizing it even when the contact pressure has fallen below 200~MPa, which is where ice undergoes a phase transition to form water at $-10^\circ$C.
Nonetheless, simulated ice withstands indentation pressures of up to 300~MPa over extended time periods, when the counterface is flat, i.e., its radius of curvature $R_\textrm{c}$ is formally infinite.
This indicates that the absence of stress gradients causing non-affine displacements impedes amorphization, which might explain why Fig.~\ref{fig:indentation}(f) reveals a similar $v_\perp$-dependent $p_\textrm{H}$ as that observed experimentally~\cite{Liefferink2021PRX}, albeit shifted to larger $v_\perp$. 
The experimental $R_\textrm{c}$ were more than four orders of magnitude larger than that of our corrugated counterface, $R_\textrm{c} \approx 12.2$~nm.
Further evidence for a scale-dependent $p_\textrm{H}$ comes from atomic-force microscopy.
Using tips with $R_\textrm{c} \approx 50$~nm, Butt {\it et al.}~\cite{Butt2000} found $p_\textrm{H}$ to be 10 times the macroscopic indentation hardness, that is, before they reinterpreted their data and potential errors to reduce the gap between results and expectations to a factor of 2.5.

When sliding at $v_0=5$~m/s under $p_\textrm{n}=20$ MPa, see Fig.~\ref{fig:indentation}(e) and movie S3, friction is lowest shortly before the tip sinks into the indentation mark.
The friction coefficient is close to its maximum value of $\mu_\textrm{max} \approx 0.5$ even before it reaches the bottom of the mark. 
This value is similar to results obtained with an atomic-force microscope and a tip radius of 200~nm~\cite{Bluhm2002JPCM}. 
In our case, the time dependence of $\mu$ is a consequence of capillary rather than mechanical effects: even a thin water layer attempts to reduce its surface energy when rough solids slide past each other.
Dynamics on the first and second stroke (periodic-boundary conditions mimic pin-on-disk tribometers) resemble each other. 
However, spatial variations in the friction decrease with each pass. 

A hydrophobic counterface behaves like the hydrophilic one, although the stiction peak and the kinetic friction are halved.
The friction now assumes the small values that would usually be associated with slippery ice, i.e., well below 0.1 at a safe distance from the indentation mark. 
Just before the downhill motion $\mu$ even becomes formally negative.

The smaller friction of curved, hydrophobic surfaces not only originates from the finite slip length of water~\cite{Baran2022PNAS}, but also because the temporospatial stress fluctuations, and thus dissipation, are diminished. 
The friction reduction is substantial although the structural differences between hydrophilic and hydrophobic setups are subtle, see Fig.~\ref{fig:indentation} (c,d).
The increase of the mean friction coefficient from a flat to a curved hydrophilic counterbody is 0.175 (from 0.103 to 0.278) versus 0.088 (from 0.015 to 0.103) in the hydrophobic case.
This leaves a missing 50\% difference of 0.175-0.088 = 0.087, which can only originate from the adhesion-enhanced viscoelastic dissipation caused by the hydrophilic surface near the leading and the trailing contact edges.
Thus, in addition to interfacial water, counterbodies must be smooth and hydrophobic for ice to have very low friction coefficients, i.e., capillary effects have to be small. 
In fact, even randomly rough hydrophobic walls can have friction coefficients of 0.02, as shown in Fig.~\ref{fig:rebuttal_graph}.
While this aligns with the low ice friction of hydrophobic surfaces, attributing the agreement solely to the underlying mechanisms would be premature. 

Using the local load-bearing ability of 300~MPa deduced from the above simulations and combining it with the geometric mean of the rough upper (30~MPa) and lower (10~MPa) bounds for the steady-state shear stress at 5~m/s and $-10^\circ$C of flat interfaces, see Fig.~\ref{fig:non_stationary}(d),  we obtain a crude lower bound for the kinetic friction coefficient of ice-on-ice of 0.05 for these conditions. 
This number is not expected to generalize to other counterfaces, if their work of adhesion is lower than that of a self-mated ice contact, in which case the counterbody may easily slide past ice or its lubrication layer. 
Therefore, friction coefficients around 0.01, as measured for steel sliding on ice at high velocities~\cite{Scherge2018L}, do not contradict our findings.

Our simulations reveal that both recent and longstanding viewpoints on the dynamics of sliding ice fail to withstand scrutiny at the molecular level.
Firstly, thin, pre-molten ice layers reduce shear stresses noticeably only during the first few nanometers of sliding, because sliding-induced amorphous zones quickly become substantially thicker than pre-existing equilibrium layers.
Their indisputable friction-reducing effect established at high humidity even for solids as hard as steel~\cite{Hong2023TO}, could be indirect in the case of ice, possibly by mitigating maximum stresses causing fewer asperities to break off and to turn into abrasives~\cite{Lever2017JG}.
Secondly, pressure-induced melting, though frequently dismissed as irrelevant outside of glaciology, matters whenever local roughness induces large stress gradients, which may enhance amorphization.
Thirdly, the small friction coefficients associated with tires on black-ice seem difficult to reach without substantial water slip at small velocities.
Although an extreme dependence of melt-water viscosity on the hydrophobicity of confining walls is theoretically possible~\cite{Canale2019PRX}, our and previous~\cite{Baran2022PNAS} simulations found no evidence of this effect.

Regardless of the points mentioned above, the key insights gained in this study follow from the observation that even the smoothest possible, incommensurate ice-ice interface forms \emph{local}, cold-welded sites, where lateral displacement triggers amorphization without heat or large normal pressures.
A counterface with comparable interactions and a similar but irregular structure will inevitably produce analogous dynamics,  leading to an approximate $w_\textrm{p} =  \sqrt{\alpha d}$ relationship as long as the amorphization is fast compared to recrystallization~\cite{Reichenbach2021PRL}.
The underlying molecule-by-molecule, or, depending on the system, atom-by-atom~\cite{Bhaskaran2010NNT} attrition should be a predominantly athermal process in material pairs lacking structural lubricity in ideal, incommensurate interfaces. 
While displacement-induced amorphization does not substantially decrease the required energy to produce structural defects, it circumnavigates the need to produce vibrational energy first, which ultimately benefits the load-bearing ability of the nearby ice. 
Given that the coldest ice crystals amorphasize fastest, that is, roughly six times faster at $T = 10$~K than at $T = -10^\circ$C, the difficulty of skiing at low temperatures must be attributed to the high (effective) viscosity of amorphous ice rather than to the commonly assumed lack of liquefaction at small temperature.

\textit{Acknowledgments---}The authors gratefully acknowledge the Gauss Centre for Supercomputing e.V. (www.gauss-centre.eu) for funding this project by providing computing time through the John von Neumann Institute for Computing (NIC) on the GCS Supercomputer JUWELS at Jülich Supercomputing Centre (JSC).
MHM acknowledges helpful discussions with Lars Pastewka, Bo Persson, and Chris May. 
MHM gratefully acknowledges the financial support from the German Research Foundation (DFG) through grant MU 1694/5-2.\\

\newpage

\pagebreak
\widetext
\begin{center}
\textbf{\large Supplementary material to \\ Cold self-lubrication of sliding ice}
\end{center}

\begin{center}
Achraf Atila$^1$, Sergey V. Sukhomlinov$^1$, and Martin H. Müser$^{1*}$ \\ \vspace{0.1cm}
$^1$Department of Material Science and Engineering, Saarland University, Saarbr\"{u}cken, 66123, Germany \\ \vspace{0.1cm}
$^*$martin.mueser@mx.uni-saarland.de
\end{center}

\renewcommand{\thefigure}{S\arabic{figure}}
\setcounter{figure}{0}

In this supplement, claims made in the main part of the manuscript are substantiated, and the overall conclusions are strengthened through additional numerical and theoretical analysis.
Occasionally, we provide information that we find interesting to mention but not central enough to be part of or touched upon in the main text.
The supplement is organized so that the information is provided, for the most part, in the same sequence as explicitly or implicitly referred to in the main text.
However, we start with the methods section.
 
\section{Methods}
The TIP4P/Ice potential~\cite{Abascal2005JCP1} is used.
It accurately reproduces the properties relevant to our study, including the pressure dependence of the melting point~\cite{Noya2007JPCC}, water's density anomaly~\cite{Vega2005JCP}, the pressure and temperature dependence~\cite{Baran2023JCP} of the viscosity,  the thickness of the water surface layer~\cite{Conde2008JCP}, and, most important for our purposes, the structure of ball-milled ice~\cite{RosuFinsen2023Science}.

The starting configurations consist of two crystals facing each other with perfectly parallel [0001] surfaces.
They are misaligned by $30^\circ$, which is realized through a $90^\circ$ rotation. 
To approximate the $\sqrt{2}$ ratio of $a$ and $b$ lattice constants while using a square simulation cell and periodic boundary conditions, fourteen unit cells are used in the $x$ and ten unit cells in the $y$ direction (before rotation) and stretched in the $x$-direction by approximately 0.5\%, while compressing the $y$-direction by a similar amount. 
This makes one water layer, which contains $1344$ water molecules, have an area of $10.7 \times 10.7$~nm$^2$.
Each crystal is made up of $14$ unit cells containing two planes and a rigid outer plane, yielding a total number of 77,952 H$_2$O molecules in the system. 
A unit cell has the height $7.35$~{\AA}, which results in a total initial height of $21.8$~nm, which includes a separation region of $\approx 4.5$~{\AA}. 

When sliding ice against sinusoidal indenters, we use $60\times 13 \times 22$ unit cells, resulting in initial system dimensions of $46.0 \times 11.4 \times 15.5$ nm$^3$ at $T = -10^\circ$C, and a total number of 287,040 H$_2$O molecules. The counterbody has a height profile of $h(x,y) = h_0 \cos(2\pi x/\lambda$), where $\lambda = 46$~nm is the length of the simulation cell parallel to the sliding direction and $h_0 = 4$~nm. Lennard-Jones interaction sites assigned to the surface of the indenter are placed in a hexagonal lattice with a nearest-neighbor spacing of $a_0 = 2.6$~{\AA}. Interactions only take place with the oxygen atoms with a length of $\sigma = 3.5$~{\AA} and $\varepsilon = 0.15$~kcal/mol. Hydrophobicity was modeled by cutting off the interaction at the minimum of the potential at $r_\textrm{c} = \sqrt[6]{2}\sigma$ and at $r_\textrm{c} = 2.5~\sigma$ for the hydrophilic surfaces, which results in the surface energy gained upon making contact to be similar to that of ice-ice interfaces.

Sliding is imposed by moving the two (rigid) outermost layers at $v_{1,2} = \pm v/2$, resulting in a relative in-plane velocity of $v$, chosen within a 1--50~m/s range, which is relevant for many situations, in particular winter sports.  
In the ice-ice simulations, normal stresses are usually set to zero and sometimes to 20~MPa, which is close to a recent estimate for the quasi-static penetration hardness of ice at -10$^\circ$C~\cite{Liefferink2021PRX}, though differences between these two sets of simulations turn out to be minor. 
To minimize disruption, only the second outermost layers are thermostatted.
Simulations are conducted with LAMMPS~\cite{Thompson2022CPC}, the visualization with OVITO~\cite{Stukowski2009MSMSE}, and the post-analysis with in-house codes. 

The amorphization width $w$ is determined as the average of an upper and a lower bound.
Both are estimated from the number density of amorphous and hexagonal-ice molecules as identified using the Chill+ algorithm~\cite{Nguyen2014JPCB} with a cutoff of $3.4$~{\AA}. 
The upper bound is located when positioning the interface between ice and water at the value of $z$ where the number density of ``amorphous molecules'' is 0.08~\AA$^{-3}$.
As the sample was cut into slices along $z$-axis with a bin size of $0.5$~\AA, the error in the upper bound of $w$ should not exceed $1.0$~\AA.
The lower bound is located in the same fashion, however, using ``hexagonal-ice  molecules'' instead of ``amorphous molecules.'' 
At low temperature, this method gives similar results as when correlating the height difference of the two outermost layers with the width $w$.
However, the latter method gives poor results at high temperatures and fails to locate a crystal surface in the absence of a counterbody.

\section{Simulation of set-ups and models other than the default}
\begin{figure}[htb]
\begin{center}
\includegraphics[width=0.9\textwidth]{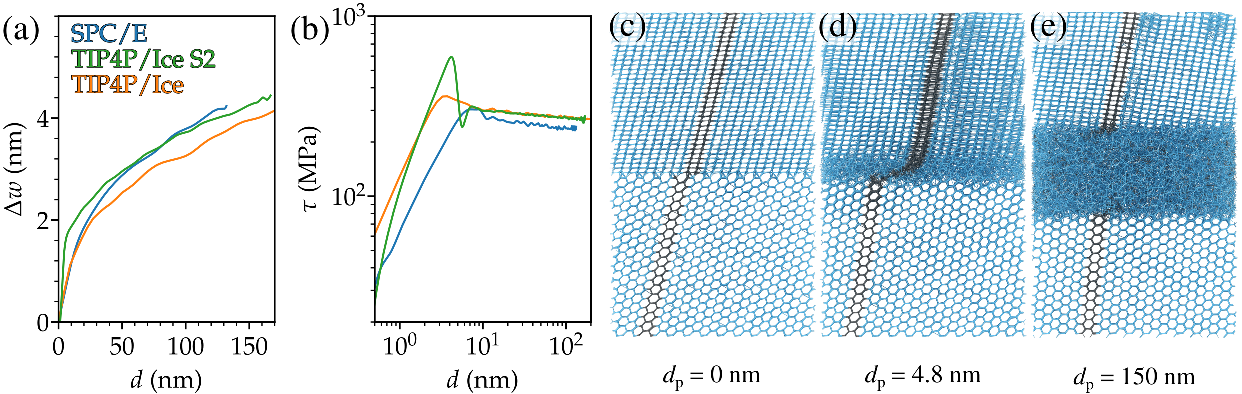}
\caption{\label{fig:alternativeModels} 
(a) Amorphization width $w$  at $T = 100$~K as a function of sliding distance $d_\textrm{p}$ for the default model (basal plane, TIP4P/Ice, orange line), the same plane but different potential (SPC/E, blue line), the default potential but the ($\overbar{1}2\overbar{1}0$) plane (TIP4P/Ice S2, green line). 
(b) Pertinent shear stress $\tau$ as a function of $w$ (same color codes as in (a)). 
(c--e) Snapshots of part of the {($\overbar{1}2\overbar{1}0$)} system at different sliding distances $d_\textrm{p}$. The slid distance in panel (e) is roughly ten times the in-plane system size. Some selected atoms representing rows orthogonal to an unstrained interface are colored in gray.}
\end{center}
\end{figure}

\begin{figure}[hbtp]
\begin{center}
\includegraphics[width=0.90\textwidth]
{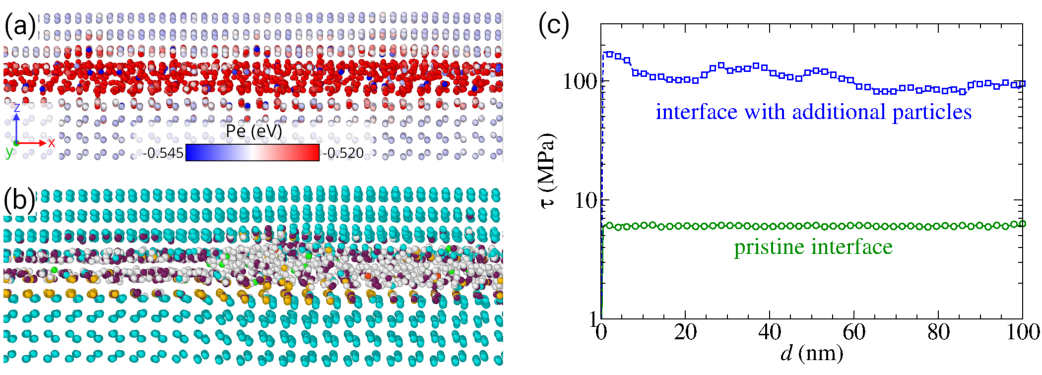}
\caption{\label{fig:mw_sliding}
The same ice-ice interface as in Fig.~1 but modeled with the mW potential and with 200 additional mW particles placed between ideal surfaces at random positions (a) after contact formation and (b) after sliding 100~nm. (c) Shear stress as a function of the sliding distance for an ideal interface (green circles) and with added particles (blue squares). 
The color coding in (a) refers to the potential energy, while that in (b) depicts local structural motifs, with local ice Ih order being depicted in cyan. 
}
\end{center}
\end{figure}

At the beginning of the main manuscript, similar observations are reported to have been made for alternative interface geometries and/or other potentials as for the default set-up.
It is beyond the scope of a supplement to summarize all of these results. 
However, Fig.~\ref{fig:alternativeModels} shows exemplarily (a) the dependence of the amorphization width $w$ on slid distance after the onset of amorphization $d_\textrm{p}$ and (b) of the shear stress $\tau$ as a function of $w$ using the same geometry but a different potential, namely the SPC/E potential~\cite{Berendsen1987JPC} and the default potential but a different geometry, in which the base planes are replaced with less densely packed {secondary-prism} planes, which are again misaligned by means of a 90$^\circ$ rotation. 

The amorphization proceeds similarly for SPC/E as for TIP4P/Ice, i.e., the $w \propto \sqrt{d_\textrm{p}}$ relationship is revealed without further ado as well as a rather minor 5\% smaller amorphization rate for SPC/E. 
Shear stresses are also of a similar order of magnitude, albeit here, the reduction is 20\%.
Given that SPC/E frequently produces ice properties with up to $\mathcal{O}(10\%)$ greater deviation from the experiment than TIP4P/Ice, it may be concluded that the observed scaling laws and order-of-magnitude estimates produced by TIP4P/Ice are reliable. 
In contrast to the model variation just discussed, using a different base plane while keeping the default potential seems to violate $w \propto \sqrt{d_\textrm{p}}$ at first sight.
Nonetheless, the $\tau(w)$ relation is clearly within 4\% of the default model. 

The reason for the different initial response of the {($\overbar{1}2\overbar{1}0$)} lies in their small packing fraction, which more readily produces cold-welded contact points than dense surfaces. 
This elevates the difference between the break-lose shear stress and the shear stress right after the depinning substantially more than for the densely packed planes, thereby corroborating the conjecture that less densely packed surfaces are less superlubric. 
As a consequence, significantly more elastic energy stored in the crystal is converted into configurational energy during depinning for the open surface than for the dense surface. 
Loosely speaking, the break-lose process of a dense surface can be seen in analogy to a second-order or rather weak first-order phase transition, while that of a loosely packed surface is clearly first order.

If modeling water with the mW model, the just-described instabilities are suppressed.
Even more, the relative sliding of incommensurate interfaces does not progressively disorder the ice crystals. 
This is believed to happen for two reasons.
First, there are almost no low-energy, interfacial particles at any point in time, which would promote local cold-welding.
Second, recrystallization is extremely fast despite the temperature being as low as $T = 10$~K.
Both effects are favored by the absence of internal degrees of freedom in mW particles, which, in the case of real water, are orientational in nature and electronic in elemental systems like silicon and diamond. 

\begin{figure}
\includegraphics[width=0.90\textwidth]{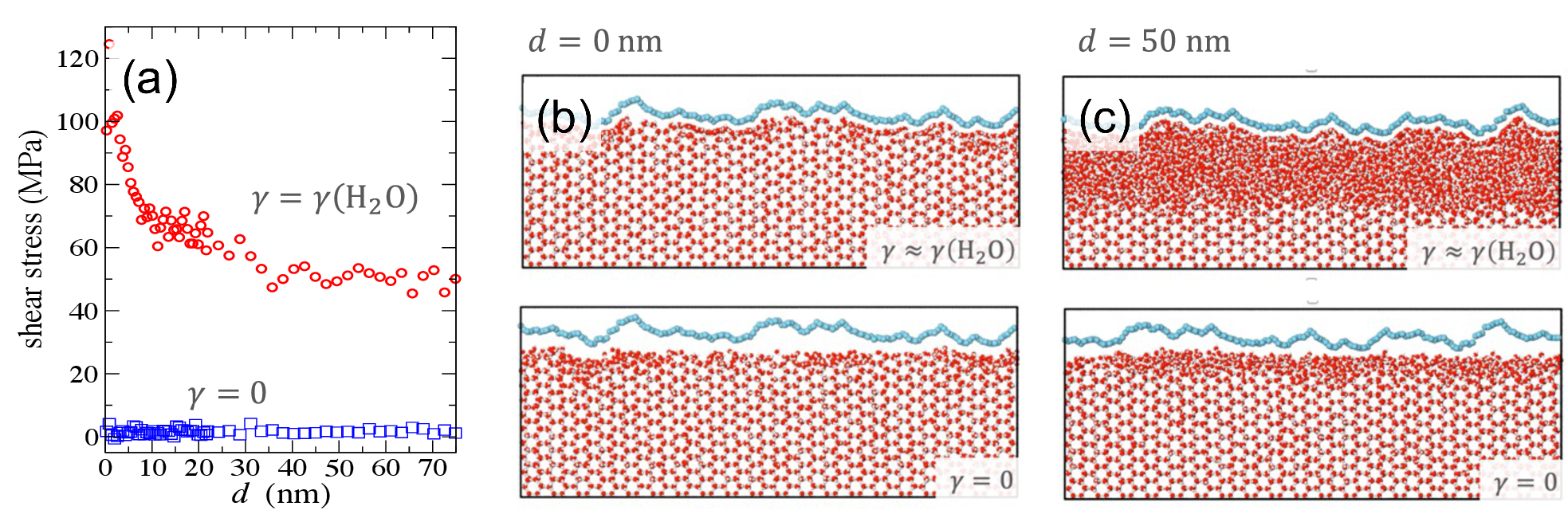}
    \caption{\label{fig:rebuttal_graph}
    (a) Shear stress as a function of sliding velocity at -10$^\circ$C and 20~MPa for hydrophilic (red) and hydrophobic (blue symbols). Snapshots during the early (b) and late (c) stages of the sliding process. Hydrophilic and hydrophobic indenters are in the top and bottom panels, respectively.
    }
\end{figure}

In the initial stages of our research, we also conducted small-scale simulations of a rigid, randomly rough surface and an ice surface at $T = -10^\circ$C and a nominal normal pressure of $p = 20$~MPa. 
Some results and snapshots are presented in Fig.~\ref{fig:rebuttal_graph} and referred to in the final discussion in the main text.
The interaction potential between indenter atoms and water molecules was treated in a similar way as for the simulations presented in Fig.~3 of the main manuscript, that is, in one set of simulations the Dupré surface energy of self-mated ice, $\gamma(\text{H}_2\text{O})$, was reproduced. 
In the other set, which addressed hydrophobic contacts, the same potential was used but cut off in its minimum and shifted such that $\gamma = 0$. 
The ice surface had not been equilibrated long enough for the spontaneous disordered film to form, yet, friction below the hydrophobic indenter and the ice friction was low from the start, yielding a small friction coefficient of roughly 0.03. 
The shear stress in the hydrophilic setup is almost two orders of magnitude in excess of that in the hydrophilic setup, despite the presence of a thick lubrication layer. 

\section{Resemblance of shear-amorphized ice with supercooled water}

\begin{figure}[hbtp]
\begin{center}
\includegraphics[width=0.425\textwidth]{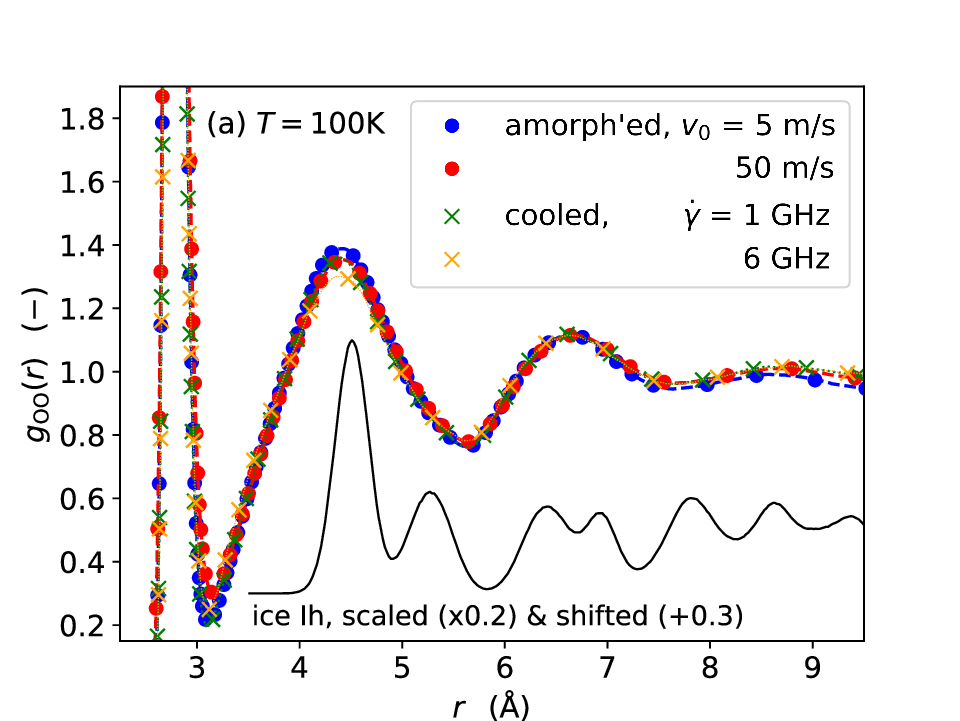}
\includegraphics[width=0.425\textwidth]{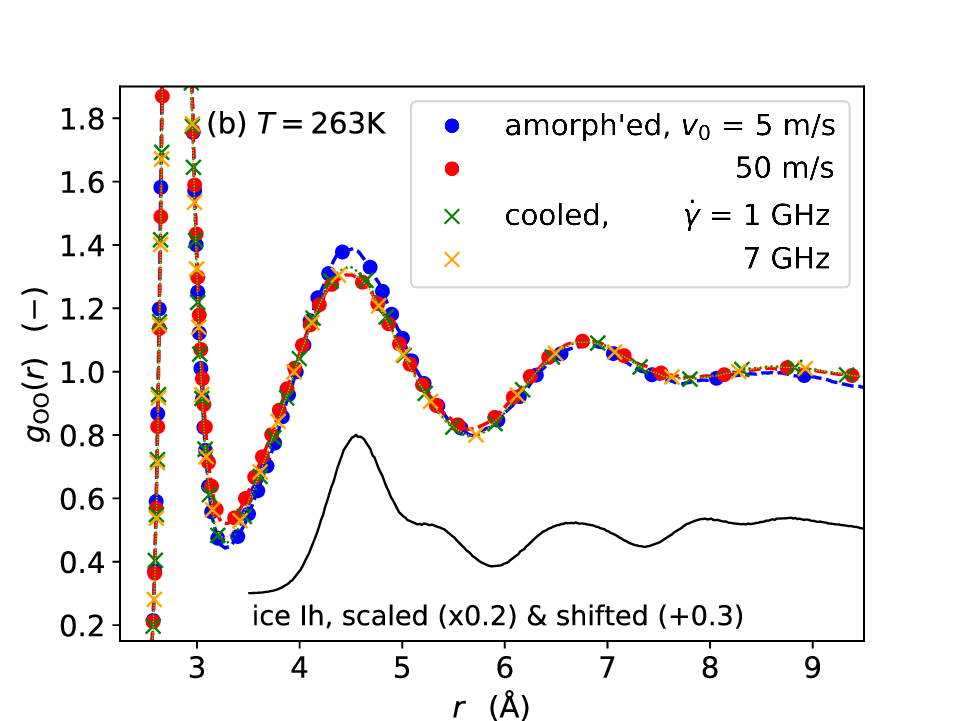}
\caption{ \label{fig:g2Combined}
The radial oxygen-oxygen distribution function $g_\textrm{OO}(r)$ in various ice and water structures at (a) $T = 100$~K and (b) $T = 263$~K.
Data for shear-amorphized ice is drawn in circles, while that obtained by quenching water during shear is depicted by crosses.
Blue and red symbols relate to high and low shear velocities, while green and yellow correspond to high and low shear rates, respectively. 
$g_2(r)$ is also shown for hexagonal ice (gray), however, the data is shifted and rescaled. 
}
\end{center}
\end{figure}

In the main manuscript, the shear-amorphized ice is said to resemble sheared supercooled water.
First evidence for this finding is presented in Fig.~\ref{fig:g2Combined}.
It depicts the oxygen-oxygen radial distribution function $g(r)$ obtained for crystalline ice, shear-amorphized ice, and sheared bulk water quenched instantaneously from ambient temperature to 100~K and 263~K in panels (a) and (b), respectively. 
While the various $g(r)$ have minor differences in the (cut-off and not fully shown) nearest-neighbor peak, deviations between peaks associated with the second and third peak are minor and reminiscent of the damped oscillations as they are typical for dense, equilibrium liquids. 
Not surprisingly, increasing shear rate or velocity reduces the heights of the peaks, however, only marginally.
In large $r$, the differences between shear-amorphized ice and sheared, supercooled water are more noticeable.
This observation can probably be rationalized by the fact that the shear-amorphized ice is effectively confined, whereby the nature of the Ornstein-Zernicke oscillations must show different asymptotic behavior.
However, this is speculative, beyond the focus of this work, and to be investigated further. 
Not only changing shear rates/velocity has a minor effect on $g(r)$, but also the 163~K temperature increase from 100 to 263~K, which is not necessarily expected.  
Crystalline ice exhibits noticeably different behavior in $g(r)$ than sheared, disordered ice, i.e., relatively sharp peaks in the radial distribution at 100~K, which are substantially smeared out at $263$~K.

\begin{figure}[hbtp]
\begin{center}
\includegraphics[width=0.85\textwidth]{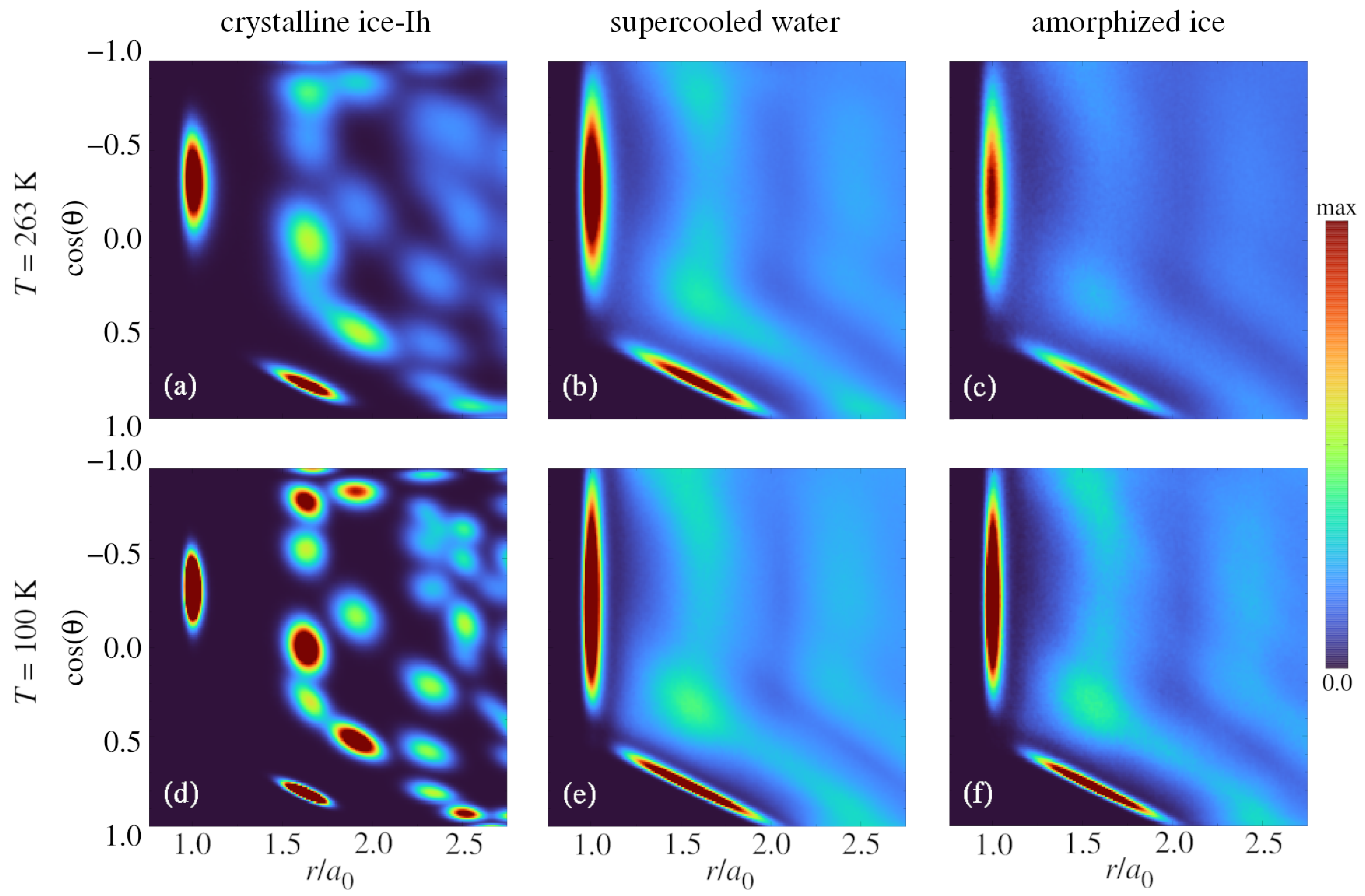}
\caption{ \label{fig:g3Combined}
The oxygen-oxygen-oxygen mixed radial, angular distribution function $g_3(r, \cos\theta)$ of (a) hexagonal ice, (b) amorphous layer obtained while sliding with the speed of 5~m/s, and (c) water, all at 263~K. (d) same as (a), (e) same as (b), and (f) same as (c), but all at 100~K.
}
\end{center}
\end{figure}

Since the radial distribution does not allow one to deduce bond angles with high reliability and bond angles might differ between shear-amorphized and sheared supercooled water, Fig.~\ref{fig:g3Combined} presents results on the recently introduced mixed radial, angular distribution function $g_3(r,\cos\theta)$\cite{Sukhomlinov2020JCP}.
It is proportional to the probability density of three atoms $i$, $j$, and $k$ to adopt a bond angle $\theta_{ijk}$ on atom $j$, when $i$ and $j$ are nearest neighbors while atoms $j$ and $k$ are separated by a distance $r$.
For example, $g_3(r,\cos\vartheta)$ shows a peak near $\cos\vartheta = 1/3$ in a tetrahedral network at a nearest-neighbor distance $r$.
Although $g_3(r,\cos\theta)$ clearly reveals orientational order in crystalline ice beyond nearest neighbors, there is no similarly strong order in the sheared waters.
Differences between different temperatures and shear-amorphized ice vs. sheared, supercooled water are again minor.

\begin{figure*}[hbtp]
\begin{center}
\includegraphics[width=0.8\textwidth]{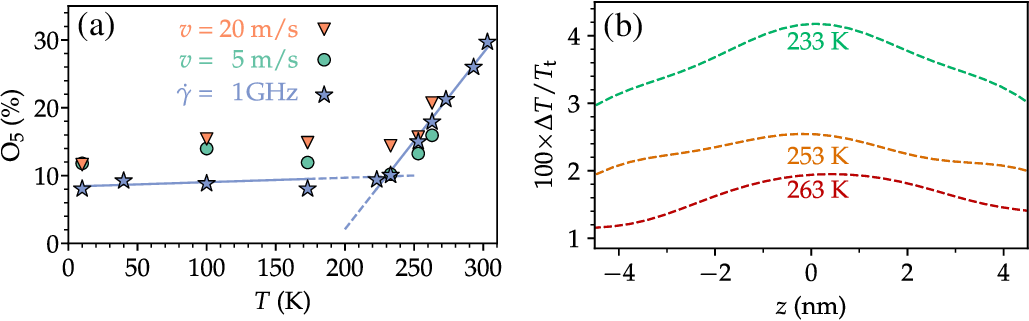}
\caption{\label{fig:FFTemp}
(a) Concentration of five-coordinated oxygen atoms in shear-amorphized ice (red circles for $v = 5$~m/s, green triangles for $v = 20$~m/s), and for sheared bulk water supercooled from 273~K for $\dot{\gamma}$ = 1 GHz (blue stars).
Light blue solid lines show linear fits to the $\dot{\gamma}=1$~GHz data. Lines are continued by dashed lines outside the domain, where lines were fitted. 
(b) Relative temperature changes of shear-amorphized ice at different target temperatures $T_\textrm{t}$ for different positions along the $z$-axis.   
The sliding velocity was 5~m/s in all cases, and the measurement was performed each time for {$w \approx 1.4$~nm}. 
}
\end{center}
\end{figure*}

An important structural characteristic responsible
for the density anomaly of water and its negative pressure-viscosity coefficient is the relative number or concentration of five-coordinated water molecules $O_5$. 
Fig.~\ref{fig:FFTemp} shows that the temperature dependence of $O_5$ is again similar for shear-amorphized ice and sheared, supercooled water, even if differences between different curves are now clearly visible to the naked eye.  

An interesting feature of the $O_5(T)$ dependence, which we believe to have remained unnoticed hitherto, is that $\partial O_5(T)/\partial T$ is close to zero below $T = 230$~K but clearly positive at higher temperature. 
The slope change occurs at a temperature where the transition from so-called high-density to low-density supercooled water is believed to take place upon heating~\cite{Gallo2016CR} at ambient pressure. 
Given that five-coordinated oxygen is responsible for the density anomalies of ice and water near $T_\textrm{m}$, it is surprising that this quantity seems poorly investigated. 
Nonetheless, we abstain from discussing five-fold coordination further since the focus of this work is on ice tribology rather than on (smeared-out, non-equilibrium) phase transformations in supercooled water or shear-liquefied ice.  

\section{Sliding-induced interfacial temperature changes}

One central aspect of our work is the realization that the ordered-disordered ice interface does not substantially heat during amorphization and that heating is more significant at low temperatures than at high temperatures. 
These claims were made on temperature profiles such as those shown in Fig.~\ref{fig:FFTemp}(b).
A mean kinetic energy per degree of freedom in a bin $\left\langle T_\textrm{kin}^\textrm{DOF}\right\rangle$ is converted through the equipartition theorem to a temperature of $T = 2 \left\langle T_\textrm{kin}^\textrm{DOF}\right\rangle/k_\textrm{B}$, where $k_\textrm{B}$ is the Boltzmann constant. 

The temperature increase for $T = 263$~K at $v_0 = 20$~m/s is 1.8\%, which translates into a little less than a heating by 5~K. 
At $v_0 = 5$~m/s, the temperature increase reduces to 1.45\% and the profile is more antisymmetric than at 20~m/s (not shown).
We relate the asymmetry to the strain in the ice crystals, which is compressive (and thus stiffer) in the sliding direction and tensile (and thus softer) in the transverse direction for one crystal and the other way around for the other crystal.
This makes the crystal that is soft parallel to sliding amorphize a little more quickly than the counterbody. 

Our simulations certainly risk underestimating interfacial heating. 
Although the sliding distances studied are large by current MD standards, they remain tiny compared to experimental scales. 
More importantly, our thermostat operates only a distance of $d_\text{t} \approx 10$~nm away from the water-ice interface. 
Assuming a steady-state situation, the temperature difference between the thermostatted regions and the interface is $\Delta T = \frac{\tau v d_\text{t}}{2 \lambda}$, with $\lambda \approx 2$ W/m·K being the thermal conductivity of real and TIP4P/Ice ice near -10°C~\cite{IriarteCarretero2018PCCP}. 
Given a shear stress of $\tau = 40$ MPa at a sliding velocity of $v = 20$ m/s, a numerical value of $\Delta T = 2$ K is obtained, which is consistent with values deduced from the kinetic energy. 
Keeping the numerical values for $\tau$ and $v$, the melting temperature at the ice-water interface is reached for $d_\text{t} \approx 50$ nm. 
However, as $T$ increases, the (effective) viscosity decreases. 
More importantly, as $T$ rises toward $T_\text{m}$, the recrystallization rate drops substantially, causing $w_\text{p}$ to increase and $\tau$ to decrease.
These two effects create a strong negative feedback on heating. 
Additionally, periodic boundary conditions prevent heated amorphous ice from being squeezed out of the contact zone, helping mitigate the risk of underestimating temperature in the simulations. 
Finally, it should be noted that a (static) temperature undulation with wavelength $\Lambda$, corresponding to the contact-patch diameter, penetrate roughly a distance of $\Lambda/2\pi$ into the material due to the mathematical nature of the heat/Laplace equation. 
Thus, relatively large contact-patch radii are required to reach melting-point conditions. These seem difficult to achieve when the high penetration hardness at large sliding velocities results in small contact patches.

\section{Rough estimate of amorphization coefficients}

\begin{figure*}[h!]
\centering
\includegraphics[width=0.5\textwidth]{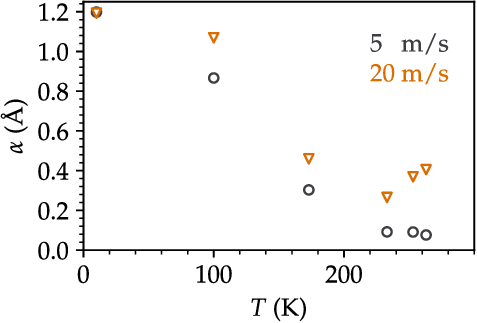}
\caption{Amorphization rate $\alpha$ as a function of the sliding temperature show for two velocities 5 m/s and 20 m/s. 
}
\label{fig:amorphizationrate}
\end{figure*}

The rate at which ice amorphasizes during the early stages of sliding is critical to the way how stress gets smaller with increasing sliding distance.
We quantify the rate with $\alpha$ as the square of the width, $w^2$, divided by the slid distance $d$.
Obtaining precise numbers for $\alpha$ at increased temperature is difficult, because there is some pre-existing molten ice during the moment of depinning. 
This causes some ambiguity both in the width of the amorphous zone and thereby on the offset of the slid distance.
Moreover, the square root scaling levels off earlier at higher temperatures so that it is only close-to-square root scaling above 233~K. 
As a consequence, the precise value of $\alpha$ depends on the length of the domain, over which $\alpha$ is determined. 
Thus, while $w(d)$ can be fitted quite easily with a $\sqrt{d}$ dependence over a larger domain, a precise value is difficult to report. 
To get consistent numbers for two different velocities, the following choices for the fitting domain of the slid distance were made:
[0:230] for 10 K, [0:100] for 100 K, [0:130] for 173 K, [0:120] for 233 K, [0:110] for 253 K, [0:200] for 263 K.
Results are reported in Fig.~\ref{fig:amorphizationrate}.

\section{Non-stationary sliding conditions}

\begin{figure*}[hbtp]
\begin{center}
\includegraphics[width=0.8\textwidth]{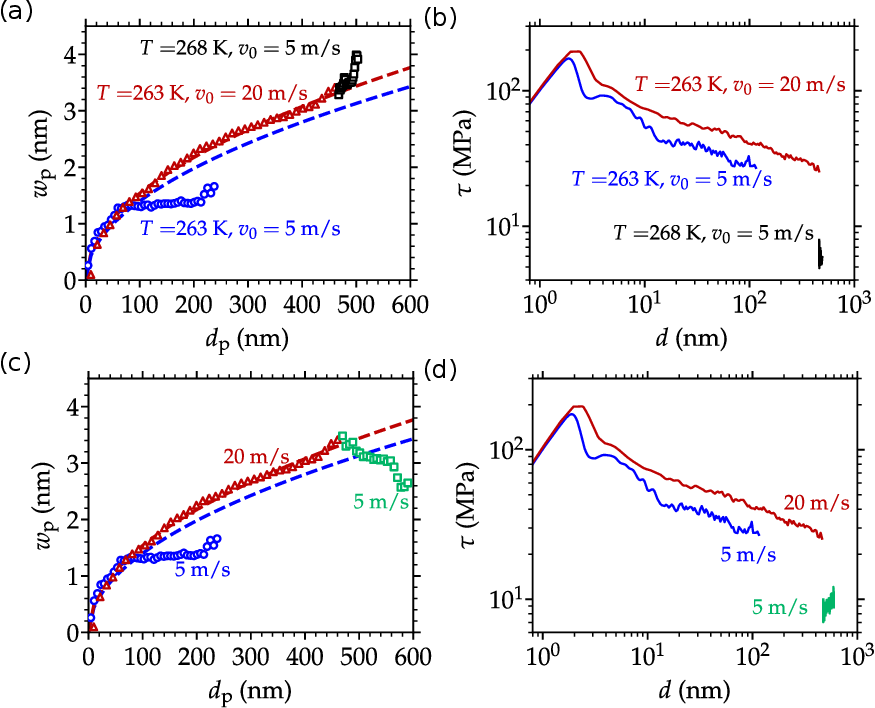}
\caption{\label{fig:non_stationary} Width $w_\text{p}$ of the liquefied zone (a, c) and shear stress (b, d) for a flat-on-flat contact, where sliding first occurred at $v = 20$~m/s and $T = -10^\circ$C for a sliding distance of $d \gtrsim 450$~nm. 
In continuation runs, the velocity was decreased to $v = 5$~m/s while maintaining the temperature (a, b) and increasing it to $T = -5^\circ$C (c, d).
}
\end{center}
\end{figure*}

In some cases, it was beneficial to deviate from the regular 'protocol' and start simulations with pre-existing, undercooled, interfacial water rather than from a hypothetical, incommensurate interface.
Results for the width of the liquefied zone and the shear stress are presented in Fig.~\ref{fig:non_stationary}.

When increasing the temperature in the follow-up run to $-5^\circ$C and reducing the velocity to $v = 5$~m/s, $w_\textrm{p}$ increased to 4~nm, while $\tau$ decreased to 6~MPa, as evidenced in  Fig.~\ref{fig:non_stationary}~(a,b).  
These numbers should be compared to those, where $T$ was kept constant, i.e., $w_\textrm{p} \approx 2.6$~nm and $\tau \approx 10$~MPa, see Fig.~\ref{fig:non_stationary}~(c,d). 
Since the product of $\tau \cdot w_\textrm{p}$ is essentially identical at the end of the two $T = -5^\circ$C and -10$^\circ$C continuation runs, the shear-stress reduction at the higher temperature is primarily an effect of a broadened amorphous zone rather than a decreased viscosity. 
The increase in $w_\text{p}$ is likely due to a reduced recrystallization rate when approaching the melting point.

Panels (c,d) of Fig.~\ref{fig:non_stationary} reveal that during the follow-up run in which only the velocity changed, the liquefied zone shrinks due to recrystallization. 
After the velocity change, the shear stress decreases instantaneously to slightly below 10~MPa directly after the velocity change, but then starts to increase again while the liquid zone shrinks in size. 
While the original $-10^\circ$C, 5~m/s simulation gives an upper bound for the steady-state shear stress of $\gtrsim 25$~MPa and a lower bound for the width of the liquid layer of $\approx 1.5$~nm, the data presented in Fig.~\ref{fig:non_stationary} provide opposite bounds: namely, 10~MPa for the shear stress and 2.6~nm for $w_\text{p}$.

\section{Indenter-induced displacement of water molecules}

\begin{figure*}[h!]
    \centering
\includegraphics[width=\textwidth]{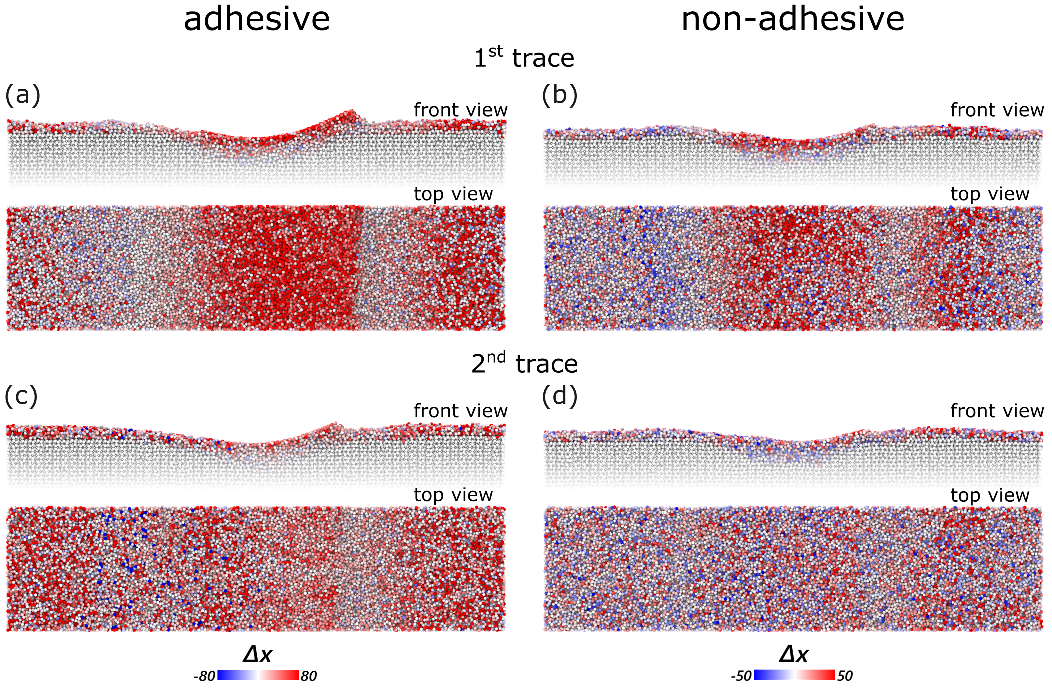}
\caption{Visualization of the displacement of water molecules for the adhesive (left) and the non-adhesive (right) indenter. The displacements for the first pass (top) and second pass (bottom) are shown separately.  }
    \label{fig:slidingindenter}
\end{figure*}

In the main text, we argue that the friction of the hydrophilic tip has a better sticking condition than for the hydrophobic tip, and claim this effect induces the larger friction of the hydrophilic tip.
In Fig.~\ref{fig:slidingindenter}, we visualize the distances by which individual molecules were displaced during the first and the second pass for each of the two indenters.
It is obvious to the naked eye that the displacement and thus the sticking was larger for the adhesive than the non-adhesive tip.
A quantitative analysis reveals that the net displaced distance, summed over all molecules, was 76~$\mu$m (first pass) and 67.5~$\mu$m (second pass) for the adhesive and 38~$\mu$m (first pass) and 31.5~$\mu$m (second pass) for the repulsive indenter.

We note that the displacement of the water in the hydrophobic case is entirely an effect arising from the roughness.
When sliding a flat hydrophilic against a flat hydrophobic wall, both being separated by a 6~nm thick and -10$^\circ$ water film, the entire water film slides with the same mean velocity of the hydrophilic surface. 

\end{document}